\documentclass{article}
\usepackage{amsmath}
\usepackage{cite}
\usepackage{graphicx}
\usepackage{dcolumn}

\begin{document}

\date{}
\title{On solutions of the Schr\"{o}dinger equation for some molecular potentials:
Power-series method}
\author{Francisco M. Fern\'{a}ndez\thanks{%
fernande@quimica.unlp.edu.ar} \\
INIFTA, DQT, Sucursal 4, C. C. 16, \\
1900 La Plata, Argentina}
\maketitle

\begin{abstract}
We show that the standard power-series method described in many textbooks of
quantum-mechanics and quantum-chemistry is simpler and more powerful than
the wavefunction approach proposed by Ikhdair and Sever [Cent. Eur. J. Phys.
6, 697 (2008)]. As illustrative examples we choose the pseudoharmonic and
Kratzer-Fues potentials already treated by those authors.
\end{abstract}

\section{Introduction}

\label{sec:intro}

Ikhdair and Sever\cite{IS08} proposed the wavefunction ansatz for the
solution of the Schr\"{o}dinger equation for some molecular potentials. In
particular, they obtained the energies and eigenfunctions in the case of the
pseudoharmonic potential and Kratzer-Fues\cite{K20,F26} one. Since the
wavefunction ansatz looks alike the standard power-series method discussed
in many textbooks on quantum mechanics\cite{CDL77} and quantum chemistry\cite
{P68}, it appears reasonable to compare both approaches.

In section~\ref{sec:models} we briefly revisit both model potentials. In
sections \ref{subsec:pseudoharmonic} and \ref{sec:Kratzer} we apply the
power-series method to the pseudoharmonic and Kratzer-Fues potentials,
respectively, and compare the results with those of Ikhdair et al. Finally,
in section~\ref{sec:conclusions} we summarize the main results and draw
conclusions.

\section{The models}

\label{sec:models}

Ikhdair and Sever presented the pseudoharmonic potential and the
Kratzer-Fues potential as $V(r)=D_{e}\left( \frac{r}{r_{e}}-\frac{r_{e}}{r}%
\right) ^{2}$ and $V(r)=D_{e}\left( 1-\frac{r_{e}}{r}\right) ^{2}$,
respectively. According to the authors ``$D_{e}$ is the dissociation energy
between two atoms in a solid and $r_{e}$ is the equilibrium intermolecular
separation''. In the first place, it is not clear what is the meaning of the
dissociation energy between two atoms in a solid. At best we may consider
two atoms in a free diatomic molecule which is what Kratzer\cite{K20} and
Fues\cite{F20} already did. In the second place, the term equilibrium
internuclear distance $r_{e}$ would be more appropriate. In the third place,
the pseudoharmonic potential does not exhibit a finite dissociation energy
because $V(r\rightarrow \infty )=\infty $. It is well known that the
dissociation energy is defined as $D_{e}=V(r\rightarrow \infty )-V\left(
r_{e}\right) $\cite{H50}.

Ikhdair and Sever considered the radial eigenvalue equation
\begin{equation}
\left\{ \frac{d^{2}}{dr^{2}}+\frac{D-1}{r}\frac{d}{dr}-\frac{\ell (\ell +D-2)%
}{r^{2}}+\frac{2\mu }{\hbar ^{2}}\left[ E-V(r)\right] \right\} \psi (r)=0,
\label{eq:Schro_IS}
\end{equation}
where, according to the authors, $\ell $ denotes the angular momentum
quantum number and $\mu $ is the reduced mass. After mentioning two atoms in
a solid and an intermolecular separation it is not clear what kind of
reduced mass is denoted by $\mu $. We assume that it is the reduced mass of
the nuclei in a diatomic molecule and that $V(r)$ describes the interaction
between the nuclei of a free molecule after the application of the
Born-Oppenheimer approximation\cite{BH54}.

In what follows we apply the textbook power-series method to the two models
discussed by Ikdahir and Sever.

\section{Pseudoharmonic oscillator}

\label{subsec:pseudoharmonic}

Ikhdair and Sever wrote the potential-energy function for the pseudoharmonic
oscillator as
\begin{equation}
V(r)=ar^{2}+\frac{b}{r^{2}}+c.  \label{eq:V_PH}
\end{equation}
The eigenvalues $E$ of the Hamiltonian operator satisfy $E(a,b,c)=E(a,b,0)+c$
so that the transition frequencies for $c\neq 0$ and $c=0$ are identical as
follows from $E_{nl}(a,b,c)-E_{n^{\prime }l^{\prime
}}(a,b,c)=E_{nl}(a,b,0)-E_{n^{\prime }l^{\prime }}(a,b,0)$. It is thus clear
that we can choose $c=0$ without altering any observable physical feature of
the system. Besides, we can transform the Schr\"{o}dinger equation into a
dimensionless eigenvalue equation by means of a simple and well known
procedure discussed elsewhere in detail\cite{F20}. In particular, we can
easily prove that $E(a,b)=\hbar \sqrt{\frac{2a}{\mu }}E\left( \frac{1}{2},%
\frac{b\mu }{\hbar ^{2}}\right) $. As a result, the radial part of the
dimensionless Schr\"{o}dinger equation can be written
\begin{equation}
\left\{ \frac{d^{2}}{dr^{2}}+\frac{\beta }{r}\frac{d}{dr}+2\left[ E-\frac{%
r^{2}}{2}-\frac{\gamma }{r^{2}}\right] \right\} R(r)=0.  \label{eq:Schro_PH}
\end{equation}
If we give suitable values to $\beta $ and $\gamma $ we obtain the
equation discussed by Ikhdair and Sever. However, we can solve a
more general problem by assuming that $\beta $ and $\gamma $ are
real positive numbers.

If we write the solution as
\begin{equation}
R(r)=r^{s}e^{-r^{2}/2}p(r),\;p(r)=\sum_{j=0}^{\infty }c_{j}r^{2j},\;s=\frac{%
\sqrt{\beta ^{2}-2\beta +8\gamma +1}-\beta +1}{2},  \label{eq:R(r)_PH}
\end{equation}
then the coefficients $c_{j}$ satisfy the two-term recurrence relation
\begin{equation}
c_{j+1}=\frac{\beta -2E+4j+2s+1}{2\left( \beta +2j+2s+1\right) \left(
j+1\right) }c_{j},\;j=0,1,\ldots .  \label{eq:Rec_rel_PH_1}
\end{equation}
Up to this point the power-series approach is identical to the wavefunction
ansatz, except for the fact that Ikhdair and Sever treated the two-term
recurrence relation as a three-term recurrence relation which makes the
calculation unnecessarily obscure and more complicated.

We can easily reduce the infinite series $p(r)$ to a polynomial of degree $n$
by requiring that $c_{n}\neq 0$ and $c_{n+1}=0$, from which we obtain
\begin{equation}
E_{n}=\frac{4n+\beta +2s+1}{2},\;n=0,1,\ldots  \label{eq:E_n_PH}
\end{equation}
that is a dimensionless version of the results $E_{p}^{\delta }$, $p=0,1$,
of Ikhdair et al. (with $\hbar =\mu =2a=1$). The recurrence relation now
becomes
\begin{equation}
c_{j+1,n}=\frac{2\left( j-n\right) }{\left( \beta +2j+2s+1\right) \left(
j+1\right) }c_{j,n},\;j=0,1,\ldots ,n-1,  \label{eq:Rec_rel_PH_2}
\end{equation}
and the infinite series $p(r)$ reduces to a polynomial of degree $n$.

A question arises if the only square-integrable solutions are those given by
the polynomial functions $p(r)$ of finite degree. In order to answer it we
rewrite the recurrence relation (\ref{eq:Rec_rel_PH_1}) as
\begin{equation}
\left( j+1\right) \frac{c_{j+1}}{c_{j}}=1-\frac{\beta +2E+2s+1}{2\left(
2j+\beta +2s+1\right) }.
\end{equation}
It is clear that $\frac{c_{j+1}}{c_{j}}\sim \frac{1}{j+1}$ for sufficiently
large values of $j$ which leads to $c_{j}\sim \frac{1}{j!}$. Therefore, for $%
E\neq E_{n}$ the infinite series behaves asymptotically as $p(r)\sim
e^{r^{2}}$ and $R(r)$ diverges at $r\rightarrow \infty $. The conclusion is
that the energy values $E_{n}$ given by equation (\ref{eq:E_n_PH}) are the
only allowed values of the energy consistent with square-integrable
solutions. This analysis is well known and is discussed in many books on
quantum mechanics\cite{CDL77} and quantum chemistry\cite{P68}. A more
rigorous procedure is shown elsewhere for the singular harmonic oscillator
that is a one dimensional version of the pseudoharmonic potential\cite{F21}.

It is worth comparing the standard power-series method with the wavefunction
ansatz. In the first place, the former is simpler and more straightforward
because it avoids the use of the determinant of a tri-diagonal matrix. The
power-series approach produces a straightforward recurrence relation,
equation (\ref{eq:Rec_rel_PH_2}), that yields the coefficients for any
solution:
\begin{equation}
R_{n}(r)=r^{s}e^{-r^{2}/2}p_{n}(r),\;p_{n}(r)=\sum_{j=0}^{n}c_{j,n}r^{2j}.
\label{eq:R_n_PH}
\end{equation}
On the other hand, Ikhdair and Sever failed to show a general equation for
the solutions with any value of the radial quantum number. Note that they
derived results for $p=0$ and $p=1$ (their radial quantum number) and then
argued that ``Following this way, we can generate a class of exact solutions
through setting $p=0,1,2,\ldots $ , etc.'' In the second place, the
power-series approach enables one to prove that the only acceptable
solutions are those given by the values of the energy that produce a
truncation of the infinite series.

\section{Kratzer-Fues potential}

\label{sec:Kratzer}

Ikhdair and Sever mentioned that ``The standard Kratzer potential is
modified by adding a $D_{e}$ term to the potential. A new type of this
potential is the modified Kratzer-type of molecular potential.'' The fact is
that $V(r)$ and $V(r)+D_{e}$ describe the same physical model using two
different origins for the energy of the system. One can choose the energy
origin arbitrarily as argued above.

Ikhdair and Sever wrote the Kratzer potential as
\begin{equation}
V(r)=\frac{a}{r}+\frac{b}{r^{2}}+c.  \label{eq:V_Kratzer}
\end{equation}
As in the preceding example we can choose $c=0$ without affecting the
physical features of the model. Following the adimensionalization procedure%
\cite{F20} mentioned above in this case we have $E(a,b)=\frac{\mu a^{2}}{%
\hbar ^{2}}E\left( -1,\frac{b\mu }{\hbar ^{2}}\right) $ and the
dimensionless radial equation becomes
\begin{equation}
\left\{ \frac{d^{2}}{dr^{2}}+\frac{\beta }{r}\frac{d}{dr}+2\left[ E+\frac{1}{%
r}-\frac{\gamma }{r^{2}}\right] \right\} R(r)=0.  \label{eq:Schro_Kratzer}
\end{equation}

The solution can be written as
\begin{equation}
R(r)=r^{s}e^{-\alpha r}p(r),\;p(r)=\sum_{j=0}^{\infty }c_{j}r^{j},\;s=\frac{%
\sqrt{\beta ^{2}-2\beta +8\gamma +1}-\beta +1}{2},\;\alpha =\sqrt{-2E},
\label{eq:R(r)_Kratzer}
\end{equation}
where the expansion coefficients satisfy
\begin{equation}
c_{j+1}=\frac{\alpha \left( \beta +2\left( j+s\right) \right) -2}{\left(
\beta +j+2s\right) \left( j+1\right) }c_{j},\;j=0,1,\ldots .
\label{eq:Rec_rel_K_1}
\end{equation}
The conditions $c_{n}\neq 0$, $c_{n+1}=0$, $n=0,1,\ldots $, lead to
\begin{equation}
\alpha =\alpha _{n}=\frac{2}{\beta +2\left( n+s\right) },  \label{eq:alpha_n}
\end{equation}
and
\begin{equation}
c_{j+1,n}=\frac{4\left( j-n\right) }{\left( \beta +j+2s\right) \left( \beta
+2\left( n+s\right) \right) \left( j+1\right) }c_{j,n},\;j=0,1,\ldots ,n-1.
\label{eq:Rec_rel_K_2}
\end{equation}

In order to show that these solutions are the only square-integrable ones we
rewrite (\ref{eq:Rec_rel_K_1}) as
\begin{equation}
(j+1)\frac{c_{j+1}}{c_{j}}=2\alpha -\frac{\alpha \left( \beta +2s\right) +2}{%
j+\beta +2s},
\end{equation}
that tells us that $\frac{c_{j+1}}{c_{j}}\sim \frac{2\alpha }{j+1}$ for
sufficiently large values of $j$, so that $c_{j}\sim \frac{(2\alpha )^{j}}{j!%
}$. Therefore, for $\alpha \neq \alpha _{n}$ the infinite series $p(r)$ in
the solution (\ref{eq:R(r)_Kratzer}) behaves asymptotically as $p(r)\sim
e^{2\alpha r}$ and $R(r)$ diverges at infinity as $R(r)\sim e^{\alpha r}$.

Once again we appreciate the advantage of using the textbook power-series
method instead of the wavefunction ansatz proposed by Ikhdair and Server.
These authors obtained expressions for $E_{0}^{\delta }$ ad $E_{1}^{\delta }$
and showed that they agree with a well known general formula.

\section{Conclusions}

\label{sec:conclusions}

The analysis given above shows that the standard power-series
method, described in most textbooks on quantum mechanics and
quantum chemistry, is simpler and more effective than the
wavefunction ansatz proposed by Ikhdair et al. In the first place,
the two-term recurrence relation enables one to obtain a general
analytical expression for the energy valid for all values of the
quantum number as well as a simple recurrence for the coefficients
necessary for obtaining any state of the system. On the other
hand, in the case of the wavefunction ansatz one has to calculate
the determinant of a tri-diagonal matrix for increasing values of
the matrix dimension. For this reason Ikhdair and Sever did not
show general expressions for the eigenvalues and eigenfunctions.
They simply compared their solutions for quantum numbers $p=0$ and
$p=1$ with the exact expressions already known. In the second
place, the two-term recurrence relation facilitates the analysis
of the solutions, like the proof that the only acceptable
solutions are those coming from the truncation condition.


\begin{thebibliography}{9}
\bibitem{IS08}  S. M. Ikhdair and R. Sever, Cent. Eur. J. Phys. 6, 697
(2008).

\bibitem{K20}  A. Kratzer, Z. Physik 3, 289 (1920).

\bibitem{F26}  E. Fues, Ann. Phys. 386, 281 (1926).

\bibitem{CDL77}  C. Cohen-Tannoudji, B. Diu, and F. Lalo\"{e}, Quantum
Mechanics (John Wiley \& Sons, New York, 1977).

\bibitem{P68}  F. L. Pilar, Elementary Quantum Chemistry (McGraw-Hill, New
York, 1968).

\bibitem{H50}  G. Herzberg, Molecular Spectra and Molecular Structure. I.
Spectra of Diatomic Molecules, Second ed. (Van Nostrand Reinhold, New York,
1950).

\bibitem{BH54}  M. Born and K. Huang, Dynamical Theory of Crystal Lattices
(Oxford University Press, New York, 1954).

\bibitem{F20}  F. M. Fern\'{a}ndez, Dimensionless equations in
non-relativistic quantum mechanics, arXiv:2005.05377 [quant-ph], (2020).

\bibitem{F21}  F. M. Fern\'{a}ndez, On the quantum-mechanical singular
harmonic oscillator, arXiv:2112.03693 [quant-ph], (2021).
\end{thebibliography}
\end{document}